\documentclass[12pt]{article}

\usepackage[margin=1in]{geometry}

\usepackage{setspace}

\usepackage{graphicx}

\usepackage{enumitem}

\usepackage{microtype}
\usepackage{subfigure}
\usepackage{booktabs}
\usepackage{comment}
\usepackage{wrapfig}
\usepackage{arydshln}
\usepackage{enumitem}
\usepackage{multirow}
\usepackage{url}
\usepackage{natbib}
\usepackage{authblk}

\RequirePackage[colorlinks,citecolor=blue,linkcolor=blue,urlcolor=blue,pagebackref]{hyperref}

\usepackage{amsmath}
\usepackage{amssymb}
\usepackage{mathtools}
\usepackage{amsfonts}
\usepackage{bm}
\usepackage{bbm}

\usepackage{algorithm}
\usepackage{algorithmic}

\def\eqref#1{equation~\ref{#1}}

\def\1{\bm{1}}

\DeclareMathAlphabet{\mathsfit}{\encodingdefault}{\sfdefault}{m}{sl}
\SetMathAlphabet{\mathsfit}{bold}{\encodingdefault}{\sfdefault}{bx}{n}

\def\calA{{\mathcal{A}}}

\def\calG{{\mathcal{G}}}
\def\calH{{\mathcal{H}}}
\def\calI{{\mathcal{I}}}

\def\calM{{\mathcal{M}}}

\def\calR{{\mathcal{R}}}

\def\calX{{\mathcal{X}}}

\def\calZ{{\mathcal{Z}}}

\def\bbE{{\mathbb{E}}}

\def\bbR{{\mathbb{R}}}

\newcommand{\E}{\mathbb{E}}

\DeclareMathOperator*{\argmax}{arg\,max}
\DeclareMathOperator*{\argmin}{arg\,min}

\DeclareMathOperator{\sign}{sign}

\newcommand{\p}[1]{\left(#1\right)}
\newcommand{\sqb}[1]{\left[#1\right]}

\newcommand{\bigp}[1]{\big(#1\big)}
\newcommand{\bigsqb}[1]{\big[#1\big]}

\newcommand{\Bigp}[1]{\Big(#1\Big)}

\newcommand{\Biggp}[1]{\Bigg(#1\Bigg)}

\newcommand{\Biggcb}[1]{\Bigg\{#1\Bigg\}}

\newcommand{\abs}[1]{\left|#1\right|}

\newcommand{\annot}[2]{\underbrace{#1}_{\text{#2}}}

\newcommand{\Exp}[1]{\mathbb{E}\left[#1\right]}
\newcommand{\bigExp}[1]{\mathbb{E}\big[#1\big]}

\newcommand{\BigExp}[1]{\mathbb{E}\Big[#1\Big]}
\newcommand{\BiggExp}[1]{\mathbb{E}\Bigg[#1\Bigg]}

\usepackage{amsthm}

\theoremstyle{plain}

\newtheorem{theorem}{Theorem}[section]

\usepackage[textsize=tiny]{todonotes}
\usepackage{multirow}
\usepackage{wrapfig}
\usepackage{subfigure}

\renewcommand{\eqref}[1]{(\ref{#1})}

\usepackage{xcolor}
\newcount\Comments  
\newcommand{\kibitz}[2]{\ifnum\Comments=1\textcolor{#1}{#2}\fi}

\usepackage[capitalize,noabbrev]{cleveref}

\allowdisplaybreaks

\title{Direct Debiased Machine Learning\\ via Bregman Divergence Minimization}

\author{Masahiro Kato\thanks{Email: \texttt{mkato-csecon@g.ecc.u-tokyo.ac.jp}}$\,$}

\affil{The University of Tokyo}

\date{\today}

\begin{document}

\maketitle 

\begin{abstract}
We develop a direct debiased machine learning framework comprising Neyman targeted estimation and generalized Riesz regression. Our framework unifies Riesz regression for automatic debiased machine learning, covariate balancing, targeted maximum likelihood estimation (TMLE), and density-ratio estimation. In many problems involving causal effects or structural models, the parameters of interest depend on regression functions. Plugging regression functions estimated by machine learning methods into the identifying equations can yield poor performance because of first-stage bias. To reduce such bias, debiased machine learning employs Neyman orthogonal estimating equations. Debiased machine learning typically requires estimation of the Riesz representer and the regression function. For this problem, we develop a direct debiased machine learning framework with an end-to-end algorithm. We formulate estimation of the nuisance parameters, the regression function and the Riesz representer, as minimizing the discrepancy between Neyman orthogonal scores computed with known and unknown nuisance parameters, which we refer to as Neyman targeted estimation. Neyman targeted estimation includes Riesz representer estimation, and we measure discrepancies using the Bregman divergence. The Bregman divergence encompasses various loss functions as special cases, where the squared loss yields Riesz regression and the Kullback-Leibler divergence yields entropy balancing. We refer to this Riesz representer estimation as generalized Riesz regression. Neyman targeted estimation also yields TMLE as a special case for regression function estimation. Furthermore, for specific pairs of models and Riesz representer estimation methods, we can automatically obtain the covariate balancing property without explicitly solving the covariate balancing objective.
\end{abstract}

\section{Introduction}
This study considers parameters of interest that depend on regression functions, such as treatment effects and policy effects. Empirical analysis often employs machine learning methods to estimate regression functions and then plugs the estimates into the estimating equations for the parameters of interest. However, while machine learning methods are effective in important situations, such as regression with high-dimensional covariates and complex regression functions, they often yield bias in the estimating equations, which prevents us from guaranteeing root-n convergence for the resulting estimator of the parameter of interest. 

To reduce this bias, debiased machine learning methods have been investigated \citep{Chernozhukov2018doubledebiased}. In debiased machine learning, we utilize Neyman orthogonal estimating equations, under which we can asymptotically eliminate the bias caused by regression function estimation with cross-fitting and mild convergence conditions for the regression function estimators. \citet{Chernozhukov2022automaticdebiased} develops the automatic debiased machine learning framework, which allows us to define and estimate the Riesz representer without specifying a particular functional form. 

The Neyman orthogonal scores usually require estimation of the Riesz representer to debias initial estimates of the regression functions. Therefore, we need to estimate the nuisance parameters, the regression functions, and the Riesz representer to construct an estimator of the parameter of interest. Although the Neyman orthogonal score mitigates the bias problem, high-quality estimation of the nuisance parameters remains important. First, even with the Neyman orthogonal estimating equation, to obtain asymptotic efficiency, we must construct nuisance parameter estimators that satisfy certain convergence rate conditions. Second, even if these conditions hold asymptotically, there is room to improve finite-sample performance. 

For example, in the treatment effect estimation literature, various methods have been proposed to improve the accuracy of treatment effect estimation. In treatment effect estimation, the Riesz representer usually depends on the propensity score, the probability of assigning treatment. Such a probability is often estimated via a logistic regression model with maximum likelihood estimation. \citet{Imai2013estimatingtreatment} proposes covariate balancing propensity scores, which fit the logistic regression model to match the first moment of the covariates between treated and control groups. \citet{Hainmueller2012entropybalancing} instead proposes not using such an explicit model and estimates weight parameters under covariate balancing conditions. These covariate balancing methods have been further extended by subsequent studies \citep{Zhao2019covariatebalancing}. From the debiased machine learning literature, \citet{Chernozhukov2024automaticdebiased} establishes Riesz regression, which aims to directly estimate the Riesz representer. Note that \citet{BrunsSmith2025augmentedbalancing} shows that Riesz regression can be derived as the dual of stable matching by \citet{Zubizarreta2015stableweights}, and we also show that the formulation is identical to least-squares importance fitting (LSIF) in the density-ratio estimation literature \citep{Kanamori2009aleastsquares}. For regression function estimation, \citet{vanderLaan2011targetedlearning} proposes Super Learner and targeted maximum likelihood estimation (TMLE). Super Learner estimates the regression function, and TMLE debiases its initial regression function estimates by using the Riesz representer. 

We propose direct debiased machine learning (DDML), a novel framework that unifies existing debiased machine learning frameworks and nuisance parameter estimation methods. Our proposed framework consists of Neyman targeted estimation and generalized Riesz regression, with the covariate balancing property. We first formulate estimation of the nuisance parameters as targeted estimation of the Neyman orthogonal score with the true nuisance parameters. We refer to this as Neyman targeted estimation, which involves estimation of the regression function and the Riesz representer. We then propose generalized Riesz regression, which estimates the Riesz representer by minimizing the Bregman divergence between the true Riesz representer and its model. The Bregman divergence is defined via a convex function chosen by us, and we can derive various estimation methods as special cases. For example, if we use the squared loss in the Bregman divergence, we obtain Riesz regression proposed in \citet{Chernozhukov2024automaticdebiased}. If we choose a well-designed convex function (see Section~\ref{sec:ateest}) in treatment effect estimation, we can derive covariate balancing propensity scores, which are proven to be equal to empirical balancing \citep{Hainmueller2012entropybalancing} by \citet{Zhao2019covariatebalancing}. Through this argument, in treatment effect estimation, we point out that specific choices of the Bregman divergence guarantee the covariate balancing property, under which the Riesz representer balances covariates between the treated and control groups.

\subsection{Contents of This Study}
We summarize these important elements below:
\begin{itemize}
    \item \textbf{Neyman targeted estimation.} We formulate estimation of the nuisance parameters as an error minimization problem between the Neyman orthogonal scores with true nuisance parameters and their models, which further decomposes into error minimization problems for the regression functions and the Riesz representer. We refer to this estimation procedure as Neyman targeted estimation.
    \item \textbf{Generalized Riesz regression.} We formulate the Riesz representer estimation problem as a Bregman divergence minimization problem. Bregman divergence depends on a convex function specified by us and yields various estimation methods by changing its definition. For example, Riesz regression and covariate balancing propensity scores are special cases. We refer to this estimation procedure as generalized Riesz regression or Bregman-Riesz regression. 
    \item \textbf{Automatic covariate balancing.} Under specific choices of Riesz representer models with basis functions and the Bregman divergence, we can automatically guarantee the covariate balancing property for the basis functions used in the Riesz representer models. 
\end{itemize}
Our proposed Riesz representer estimation method is novel, not merely a generalization of existing methods. We define the objective as weighted risk minimization to improve performance. Existing methods such as Riesz regression and covariate balancing propensity scores are special cases where we use identity weights. Using the proposed method, we demonstrate concrete implementations in several applications, such as Average Treatment Effect (ATE) estimation, ATT estimation, average marginal effect estimation, and covariate shift adaptation. 

\begin{figure}[t]
    \centering
    \includegraphics[width=0.95\linewidth]{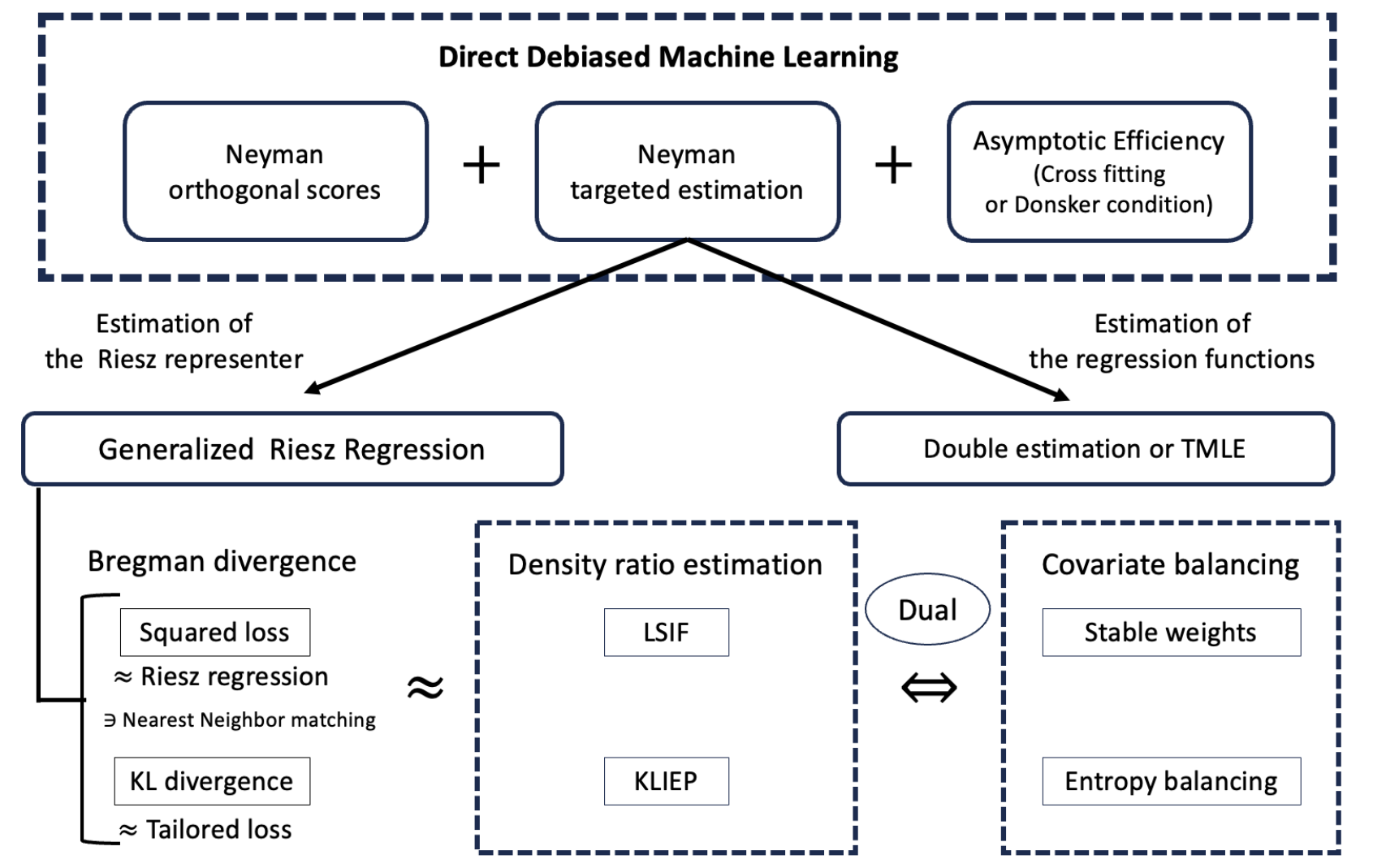}
    \caption{Concept of direct debiased machine learning.}
    \label{fig:ddml_cenpt}
\end{figure}

Furthermore, in the course of our arguments, we find the following points: 
\begin{itemize}
    \item In Neyman targeted estimation, there are several approaches to regression function estimation, which include targeted maximum likelihood estimation as a special case. 
    \item Riesz regression and direct density-ratio estimation methods coincide in several applications, such as ATE estimation and covariate shift adaptation. Riesz regression can address a wider class of problems based on its derivation via the Riesz representation theorem. In contrast, direct density-ratio estimation methods are more general in the sense of the loss function using the Bregman divergence. When using the squared loss in direct density-ratio estimation methods, direct density-ratio estimation and Riesz regression become the same in several applications. 
    \item The direct density-ratio estimation method proposed in \citet{Lin2023estimationbased} is a special case of LSIF in \citet{Kanamori2009aleastsquares} and of Riesz regression in \citet{Chernozhukov2022automaticdebiased}. For details, see \citet{Kato2025nearestneighbor}.
\end{itemize}
Thus, our DDML framework includes various existing methods, such as Riesz regression, covariate balancing propensity scores, targeted maximum likelihood estimation, density-ratio estimation, and nearest neighbor matching, as special cases. Also see Figure~\ref{fig:ddml_cenpt}.

In Section~\ref{sec:basic_idea}, we first propose a method for ATE estimation based on MSE minimization targeting the oracle score functions. In Section~\ref{sec:directdebiasedmachinelearning}, we formulate the problem more generally and establish the direct debiased machine learning framework. Direct debiased machine learning consists of two important components, the Neyman targeted estimation and generalized Riesz regression, which are introduced in Sections~\ref{sec:neymantargetedestimation} and \ref{sec:grr}, respectively. We demonstrate applications of our framework to ATE estimation, Average Treatment Effect on the Treated (ATT) estimation, Average Marginal Effect (AME) estimation, and covariate shift adaptation in Sections~\ref{sec:ateest}--\ref{sec:covshift}. In Section~\ref{sec:covariatebalancingconsistency}, we define automatic covariate balancing, which bridges Riesz regression and covariate balancing methods. 

This work is partially based on our earlier paper, \citet{Kato2025directbias}. For several theoretical and simulation study results, also see \citet{Kato2025directbias}. 

\subsection{Related Work}
Our work is mainly based on the literature on debiased machine learning, covariate balancing, TMLE, and direct density-ratio estimation. As briefly explained above and shown in the main text, these strands of literature essentially discuss the same topics from different perspectives. In this section, we introduce existing studies in these literatures. 

Efficient estimation of semiparametric models has been intensively studied in various fields, including statistics, economics, machine learning, and epidemiology. A powerful tool for discussing the optimality of semiparametric models is the asymptotic efficiency bound for regular estimators \citep{Bickel1998efficientadaptive,Vaart1998asymptoticstatistics}. It is known that efficient estimators, whose asymptotic variance matches the efficiency bound, are regular and asymptotically linear (RAL) for the semiparametric efficient influence function. To construct such efficient estimators, existing studies suggest one-step bias correction \citep{Vaart2002semiparametricstatistics}, estimating equations, or TMLE, given estimated nuisance parameters. These estimators are constructed via the efficient influence function or the Riesz representer, defined later, and under certain conditions they are shown to be efficient. The conditions usually require that the estimators of the nuisance parameters satisfy complexity and convergence rate conditions. The complexity condition is often represented by the Donsker condition, and if it is not satisfied, we typically apply sample splitting to construct the estimators \citep{Klaassen1987consistentestimation,Zheng2011crossvalidatedtargeted}.

These arguments about constructing efficient estimators have been systematized as debiased machine learning \citep{Chernozhukov2018doubledebiased,Chernozhukov2022automaticdebiased}. \citet{Chernozhukov2018doubledebiased} packages the estimating equation approach with sample splitting as Neyman orthogonal estimating equations with cross-fitting. In this framework, we first estimate the nuisance parameters using cross-fitting and plug these estimators into the Neyman orthogonal score. Then, we estimate the parameter of interest by finding a parameter such that the score is zero. The Neyman orthogonal score depends on the Riesz representer, and \citet{Chernozhukov2022automaticdebiased} shows that we can define the Neyman orthogonal estimating equation without explicitly knowing the Riesz representer. \citet{Chernozhukov2024automaticdebiased} develops Riesz regression to estimate the Riesz representer under the squared loss. Note that the Neyman orthogonal score usually corresponds to the semiparametric efficient score, and the Riesz representer corresponds to the bias-correction term used in one-step bias correction and clever covariates in TMLE. 

Semiparametric models are often used for causal inference, where the main task is to estimate treatment effects. For better estimation of treatment effects, estimation of the propensity score, the probability of assigning treatment, has been intensively studied. For example, the propensity score can be used to eliminate selection bias\footnote{The meanings of bias and debias here differ from those in debiased machine learning.}. In addition, the Riesz representer usually includes the inverse of the propensity score. Thus, better estimation of the propensity score is expected to improve treatment effect estimation. The propensity score can also be interpreted as weights for the outcomes.

Randomized controlled trials are the gold standard for causal inference, where treatments are assigned while maintaining balance between treatment groups. However, they are not always feasible, and we aim to estimate causal effects from observational data, where there often exists imbalance between treatment groups. To correct the imbalance, propensity scores or balancing weights have been proposed. 
Covariate balancing is a popular approach for propensity score or balancing weight estimation. The propensity score has been proven to be a balancing score, and based on this property, existing studies propose estimating the propensity score or the weights so that the weighted covariate moments between treated and control groups match. \citet{Imai2013estimatingtreatment} proposes estimating the propensity score by matching first moments, and \citet{Hazlett2020kernelbalancing} extends the method to higher-moment matching by mapping the covariates into a high-dimensional space with basis functions. On the other hand, estimation methods that do not directly specify the propensity score model have also been proposed. Such methods are called empirical balancing and include entropy balancing \citep{Hainmueller2012entropybalancing} and stable matching \citep{Zubizarreta2015stableweights}. These two approaches appear different but are essentially the same via a duality relationship, as proven by \citet{Zhao2019covariatebalancing} and \citet{BrunsSmith2025augmentedbalancing}. 

Our proposed framework is also inspired by the direct density-ratio estimation framework \citep{Sugiyama2012densityratio}. We refer to the ratio between two densities as the density ratio. Density-ratio estimation has attracted considerable attention as an essential task in various machine learning problems, such as regression under covariate shift \citep{Shimodaira2000improvingpredictive,Reddi2015doublyrobust,Kato2024doubledebiasedcovariateshift}, learning with noisy labels \citep{Liu2014classificationwith,Fang2020rethinkingimportance}, anomaly detection \citep{Smola2009relativenovelty,Hido2008inlierbased,Abe2019anomalydetection}, two-sample testing \citep{Keziou2005testof,Kanamori2010fdivergence,Sugiyama2011leastsquarestwosample}, change-point detection \citep{Kawahara2009changepointdetection}, causal inference \citep{Uehara2020offpolicy}, recommendation systems \citep{Togashi2021densityratiobased}. While the density ratio can be estimated by separately estimating each density, such an approach may magnify estimation errors by compounding two independent estimations. To address this issue, end-to-end, direct density-ratio estimation methods have been studied, including moment matching \citep{Huang2007correctingsample,Gretton2009covariateshift}, probabilistic classification \citep{Qin1998inferencesfor,Cheng2004semiparametricdensity}, density matching \citep{Nguyen2010estimatingdivergence}, density-ratio fitting \citep{Kanamori2009aleastsquares}, and positive–unlabeled learning \citep{Kato2019learningfrom}. It is also known that when complicated models such as neural networks are used for this task, the loss function can diverge in finite samples \citep{Kiryo2017positiveunlabeledlearning}. Therefore, density-ratio estimation methods with neural networks have been investigated \citep{Kato2021nonnegativebregman,Rhodes2020}.

\section{Basic Idea: Direct ATE Estimation with Squared Loss}
\label{sec:basic_idea}
To explain a basic idea of our DDML framework, we introduce an example in ATE estimation. Let $X = (D, Z)$, where $D \in \{1, 0\}$ denotes a binary treatment indicator, and $Z \in \calZ$ denotes covariates with covariate space $\calZ$. Let $Y = D \cdot Y(1) + (1 - D) \cdot Y(0)$, where $(Y(1), Y(0))$ are conditionally independent of treatment $D$ given covariates $Z$. We denote the pair of $Y$ and $X$ by $W = (X, Y)$. 

The parameter of interest is the ATE, defined as 
\[\tau_0 = \Exp{Y(1) - Y(0)}.\]
Our goal is to estimate $\tau_0$ by using observations $\{W_i\}^n_{i=1}$, where $W_i$ is an i.i.d. copy of $W$. 
Let $\gamma(X)\coloneqq \Exp{Y\mid X = (D,Z)}$ be the conditional expected outcome $Y$ given $X$. Let $\pi_0(Z) \coloneqq \Pr(D=1\mid Z)$ be the propensity score, the probability of assigning treatment $1$. We assume that $0 < \pi_0(Z) < 1$ almost surely. 

\subsection{ATE Estimator} 
We can estimate the ATE by using the Augmented Inverse Probability Weighting (AIPW) estimator, defined as follows:
\[\widehat{\tau}^{\text{AIPW}} \coloneqq \frac{1}{n}\sum^n_{i=1}h^{\mathrm{AIPW}}(W; \widehat{\eta}),\]
where for $\eta = (\gamma, \pi)$, $h^{\mathrm{AIPW}}(W; \eta)$ is defined as
\begin{align*}
    h^{\mathrm{AIPW}}(W; \eta) \coloneqq \p{\frac{D}{\pi(Z)} - \frac{1 - D}{1 - \pi(Z)}}\Bigp{Y - \gamma(X)} + \gamma((1, Z)) - \gamma((0, Z)),
\end{align*}
and $\widehat{\eta} = (\widehat{\gamma}, \widehat{\pi})$ is a pair of estimators of $\gamma_0$ and $\pi_0$. This estimator has various desirable theoretical properties. In particular, its asymptotic efficiency and double robustness play important roles in ATE estimation.  

In ATE estimation, the efficient score function is given as
\[\psi(W; \eta_0) = h^{\text{AIPW}}(W; \eta_0) - \tau_0,\]
where $\eta_0 \coloneqq (\gamma_0, \pi_0)$ is the pair of the true functions $\gamma_0$ and $\pi_0$. Regular and asymptotically linear (RAL) estimators based on the efficient score are known to be asymptotically efficient, that is, among regular estimators, there is no estimator whose asymptotic variance is lower than that of the regular and asymptotically linear (RAL) estimator, $\widehat{\tau}$.

\subsection{Targeting the Oracle for Estimating the Nuisance Parameters} 
In order to use the AIPW estimator in practice, we need to estimate $\eta_0$ and replace it with its estimator $\widehat{\eta}$ in the estimator. Such functions $\gamma_0$ and $\pi_0$ are called nuisance parameters. 

Define an oracle as 
\[\widetilde{\tau} \coloneqq \frac{1}{n}\sum^n_{i=1}h^{\text{AIPW}}(W_i; \eta_0).\]
Then, we decompose the error between the oracle $\tau_0$ and $\widehat{\tau}^{\text{AIPW}}$ as
\begin{align*}
    &\tau_0 - \widehat{\tau}^{\text{AIPW}}\\
    &= \tau_0 - \widetilde{\tau} + \widetilde{\tau} - \widehat{\tau}^{\text{AIPW}}\\
    &= \frac{1}{n}\sum^n_{i=1}\Bigp{\tau_0 - h^{\text{AIPW}}(W_i; \eta_0)} + \frac{1}{n}\sum^n_{i=1}\Bigp{h^{\text{AIPW}}(W_i; \eta_0) - h^{\text{AIPW}}(W_i; \widehat{\eta})},
\end{align*}
where $\tau_0 - h^{\text{AIPW}}(W_i; \eta_0)$ corresponds to the efficient score, and $h^{\text{AIPW}}(W_i; \eta_0) - h^{\text{AIPW}}(W_i; \widehat{\eta})$ is the error incurred when using estimated nuisance parameters instead of the true ones. 

A straightforward approach is to minimize the empirical mean squared error (MSE): 
\[\widehat{\eta} \coloneqq \argmin_{\eta \in \calM}\frac{1}{n}\sum^n_{i=1}\Bigp{h^{\text{AIPW}}(W_i; \eta_0) - h^{\text{AIPW}}(W_i; \eta)}^2,\]
where $\calM = \calG \times \Pi$ is a model of $\eta$, and $\calG$ and $\Pi$ are models of $\gamma_0$ and $\pi_0$. 
Since this optimization problem is infeasible because $\eta_0$ is unknown, we consider an alternative objective function.

To construct an alternative, we estimate $\eta_0$ by minimizing the population MSE:
\begin{align*}
    \eta^* = (\gamma^*, \pi^*) \coloneqq \argmin_{\eta \in \calM}\BigExp{\Bigp{h^{\mathrm{AIPW}}(W; \eta_0) - h^{\mathrm{AIPW}}(W; \eta)}^2}.
\end{align*}
Surprisingly, for this population MSE minimization, we can obtain an equivalent objective function that does not include the unknown $\eta_0$. 

\paragraph{Estimation of the propensity score.}
Given $\gamma$, we estimate $\pi_0$ by the following mean squared error minimization:
\begin{align}
\label{eq:basic_original_target}
    \pi^*(\gamma) \coloneqq \argmin_{\pi \in \Pi}\BigExp{\Bigp{h^{\mathrm{AIPW}}(W; \eta_0) - h^{\mathrm{AIPW}}(W; (\gamma, \pi))}^2},
\end{align}
where recall that $\Pi$ is a model of $\pi_0$. If $\pi_0 \in \Pi$, then $\pi^* = \pi_0$ holds and $\eta_0 = (\gamma_0, \pi_0)$. 

This MSE minimization problem can be written in a form that does not include the unknown $\pi_0$:
\begin{align}
\label{eq:basic_direct_target}
    &\pi^\dagger(\gamma) \coloneqq \argmin_{\pi \in \Pi}\Biggcb{\Exp{\p{- 2\p{\frac{1}{\pi(Z)} + \frac{1}{1 - \pi(Z)}} + \p{\frac{D}{\pi(Z)} - \frac{1 - D}{1 - \pi(Z)}}^2}\Bigp{Y - \gamma_0(X)}^2}\\
    &+ \Exp{\p{\p{1 - \frac{D}{\pi(Z)}}\Bigp{\gamma_0((1, Z)) - \gamma((1, Z))} - \p{1 - \frac{1 - D}{1 - \pi(Z)}}\Bigp{\gamma_0((0, Z)) - \gamma((0, Z))}}^2}}.\nonumber
\end{align}

\begin{theorem}
\label{thm:basic_direct}
    For \eqref{eq:basic_original_target} and \eqref{eq:basic_direct_target}, $\pi^*(\gamma) = \pi^\dagger(\gamma)$ holds. We also have
    \begin{align*}
        &\pi^*(\gamma_0) = \pi^\dagger(\gamma_0) \coloneqq\\
        &\argmin_{\pi \in \Pi}\Biggcb{\Exp{\p{- 2\p{\frac{1}{\pi(Z)} + \frac{1}{1 - \pi(Z)}} + \p{\frac{D}{\pi(Z)} - \frac{1 - D}{1 - \pi(Z)}}^2}\Bigp{Y - \gamma_0(X)}^2}}.
    \end{align*}
\end{theorem}
The proof is shown in Section~\ref{sec:proof:thm:basic_direct}. If $\pi_0 \in \Pi$, we have $\pi^\dagger(\gamma) = \pi_0$ for any $\gamma$. 

\paragraph{Estimation of the regression function.}
Given $\pi$, we aim to estimate $\gamma_0$ as
\begin{align}
\label{eq:basic_original_target2}
    \gamma^*(\pi) \coloneqq \argmin_{\gamma \in \calG}\BigExp{\Bigp{h^{\mathrm{AIPW}}(W; \eta_0) - h^{\mathrm{AIPW}}(W; (\gamma, \pi))}^2},
\end{align}
where recall that $\calG$ is a model of $\gamma_0$. If $\gamma_0 \in \calG$, then $\gamma^* = \gamma_0$ holds. 
This is equivalent to
\[\gamma^*(\pi) \coloneqq \argmin_{\gamma \in \calG}\BiggExp{\Biggp{\p{1 - \frac{D}{\pi(Z)}}\Bigp{\gamma_0((1, Z)) - \gamma((1, Z))} - \p{1 - \frac{1 - D}{1 - \pi(Z)}}\Bigp{\gamma_0((0, Z)) - \gamma((0, Z))}}^2}.\]
Unfortunately, unlike the propensity score estimation, we cannot eliminate $\gamma_0$ in the optimization. Hence, we consider minimizing an upper bound of the objective function:
\begin{align*}
    \gamma^\dagger(\pi) &\coloneqq \argmin_{\gamma \in \calG}\Biggcb{\BiggExp{\frac{D}{\pi(Z)}\p{\frac{1}{\pi(Z)} - 1}\Bigp{Y - \gamma((1, Z))}^2}\\
    &\ \ \ \ \ \ + \BiggExp{\frac{1 - D}{1 - \pi(Z)}\p{\frac{1}{1 - \pi(Z)} - 1}\Bigp{Y - \gamma((0, Z))}^2}}.
\end{align*}
Here, we derive the above upper bound as follows:
\begin{align*}
    &\BiggExp{\Biggp{\p{1 - \frac{D}{\pi(Z)}}\Bigp{\gamma_0((1, Z)) - \gamma((1, Z))} - \p{1 - \frac{1 - D}{1 - \pi(Z)}}\Bigp{\gamma_0((0, Z)) - \gamma((0, Z))}}^2}\\
    &\leq 2\BiggExp{\p{1 - \frac{D}{\pi(Z)}}^2\Bigp{\gamma_0((1, Z)) - \gamma((1, Z))}^2 + \p{1 - \frac{1 - D}{1 - \pi(Z)}}^2\Bigp{\gamma_0((0, Z)) - \gamma((0, Z))}^2}\\
    &= 2\p{\BiggExp{\p{\frac{1}{\pi(Z)} - 1}\Bigp{\gamma_0((1, Z)) - \gamma((1, Z))}^2} + \BiggExp{\p{\frac{1}{1 - \pi(Z)} - 1}\Bigp{\gamma_0((0, Z)) - \gamma((0, Z))}^2}}.
\end{align*}

\paragraph{Joint estimation of the nuisance parameters.}
Based on the above arguments, we present the first basic two-step algorithm. In the subsequent subsection, we provide an iterative algorithm as an extension of this algorithm. 

\subsection{Nuisance Parameter Estimation}
First, we estimate the nuisance parameters using the following two-step procedure:
\begin{enumerate}
    \item Estimate $\gamma_0$ as 
    \[\widehat{\gamma}^{(0)} \coloneqq \argmin_{\gamma \in \calG} \frac{1}{n}\sum^n_{i=1}\Bigp{Y_i - \gamma(X_i)}^2\]
    \item Estimate $\pi_0$ as
        \begin{align*}
            &\widehat{\pi} \coloneqq \argmin_{\pi \in \Pi}\\
            &\Biggcb{\frac{1}{n}\sum^n_{i=1}\p{- 2\p{\frac{1}{\pi(Z)} + \frac{1}{1 - \pi(Z)}} + \p{\frac{D}{\pi(Z)} - \frac{1 - D}{1 - \pi(Z)}}^2}\Bigp{Y - \widehat{\gamma}^{(0)}(X)}^2}.
        \end{align*}
\end{enumerate}

\subsection{Iterative Nuisance Parameter Estimation}
We can also estimate the nuisance parameters as
\begin{enumerate}
    \item Estimate $\gamma_0(X)$ as
    \[\widehat{\gamma}^{(0)} \coloneqq \argmin_{\gamma \in \calG} \frac{1}{n}\sum^n_{i=1}\Bigp{Y_i - \gamma(X_i)}^2\]
    \item For $t = 1, 2, \dots, T$, iterate the following procedure:
    \begin{enumerate}
        \item Estimate $\pi_0$ as
        \begin{align*}
            &\widehat{\pi}^{(t)} \coloneqq \argmin_{\pi \in \Pi}\\
            &\Biggcb{\frac{1}{n}\sum^n_{i=1}\p{- 2\p{\frac{1}{\pi(Z)} + \frac{1}{1 - \pi(Z)}} + \p{\frac{D}{\pi(Z)} - \frac{1 - D}{1 - \pi(Z)}}^2}\Bigp{Y - \widehat{\gamma}^{(t-1)}(X)}^2}.
        \end{align*}
        \item Estimate $\gamma_0$ as
        \begin{align*}
            &\widehat{\gamma}^{(t)} \coloneqq \argmin_{\gamma \in \calG} \Biggcb{\frac{1}{n}\sum^n_{i=1}\Biggp{\frac{D_i}{\widehat{\pi}^{(t)}(Z_i)}\p{\frac{1}{\widehat{\pi}^{(t)}(Z_i)} - 1}\Bigp{Y_i - \gamma((1, Z_i))}^2\\
            &\ \ \ \ \ \ \ \ \ \ \ \ \ \ \ \ \ \ + \frac{1 - D_i}{1 - \widehat{\pi}^{(t)}(Z_i)}\p{\frac{1}{1 - \widehat{\pi}^{(t)}(Z_i)} - 1}\Bigp{Y_i - \gamma((0, Z_i))}^2}}.
        \end{align*}
    \end{enumerate}
\end{enumerate}
\subsection{Derivation of Theorem~\ref{thm:basic_direct}}
\label{sec:proof:thm:basic_direct}
In this section, we derive Theorem~\ref{thm:basic_direct}; that is, the optimization problem \eqref{eq:basic_original_target} is equivalent to \eqref{eq:basic_direct_target}. While \eqref{eq:basic_original_target} includes the unknown $\pi_0$, \eqref{eq:basic_direct_target} does not include $\pi_0$. 

First, we have
\begin{align*}
    &\BigExp{\Bigp{h^{\mathrm{AIPW}}(W; \eta_0) - h^{\mathrm{AIPW}}(W; \eta)}^2}\\
    &= \BigExp{\Bigp{h^{\mathrm{AIPW}}(W; (\gamma_0, \pi_0)) - h^{\mathrm{AIPW}}(W; (\gamma, \pi))}^2}\\
    &= \BigExp{\Bigp{h^{\mathrm{AIPW}}(W; (\gamma_0, \pi_0)) - h^{\mathrm{AIPW}}(W; (\gamma_0, \pi))\\
    &\ \ \ \ \ \ \ \ \ \ \ \ \ \ \ \ \ \ \ \ \ \ \ \ \ \ \ \ \ \ + h^{\mathrm{AIPW}}(W; (\gamma_0, \pi)) - h^{\mathrm{AIPW}}(W; (\gamma, \pi))}^2}\\
    &= \BigExp{\Bigp{h^{\mathrm{AIPW}}(W; (\gamma_0, \pi_0)) - h^{\mathrm{AIPW}}(W; (\gamma_0, \pi))}^2}\\
    &\ \ \ \ \ \ \ \ \ \ \ \ \ \ \ \ \ \ \ \ \ \ \ \ \ \ \ \ \ \ + \BigExp{\Bigp{h^{\mathrm{AIPW}}(W; (\gamma_0, \pi)) - h^{\mathrm{AIPW}}(W; (\gamma, \pi))}^2}.
\end{align*}

We compute $\BigExp{\Bigp{h^{\mathrm{AIPW}}(W; (\gamma_0, \pi_0)) - h^{\mathrm{AIPW}}(W; (\gamma_0, \pi))}^2}$ as
\begin{align*}
    &\BigExp{\Bigp{h^{\mathrm{AIPW}}(W; (\gamma_0, \pi_0)) - h^{\mathrm{AIPW}}(W; (\gamma_0, \pi))}^2}\\
    &= \Exp{\p{\p{\frac{D}{\pi_0(Z)} - \frac{1 - D}{1 - \pi_0(Z)}}\Bigp{Y - \gamma_0(X)} - \p{\frac{D}{\pi(Z)} - \frac{1 - D}{1 - \pi(Z)}}\Bigp{Y - \gamma_0(X)}}^2}\\
    &= \Exp{\p{\p{\p{\frac{D}{\pi_0(Z)} - \frac{1 - D}{1 - \pi_0(Z)}} - \p{\frac{D}{\pi(Z)} - \frac{1 - D}{1 - \pi(Z)}}}\Bigp{Y - \gamma_0(X)}}^2}\\
    &= \Exp{\p{\p{\frac{D}{\pi_0(Z)} - \frac{1 - D}{1 - \pi_0(Z)}} - \p{\frac{D}{\pi(Z)} - \frac{1 - D}{1 - \pi(Z)}}}^2\Bigp{Y - \gamma_0(X)}^2}.
\end{align*}

We compute $\BigExp{\Bigp{h^{\mathrm{AIPW}}(W; (\gamma_0, \pi)) - h^{\mathrm{AIPW}}(W; (\gamma, \pi))}^2}$ as 
\begin{align*}
    &\BigExp{\Bigp{h^{\mathrm{AIPW}}(W; (\gamma_0, \pi)) - h^{\mathrm{AIPW}}(W; (\gamma, \pi))}^2}\\
    &= \BiggExp{\Biggp{\p{\frac{D}{\pi(Z)} - \frac{1 - D}{1 - \pi(Z)}}\Bigp{Y - \gamma_0(X)} + \gamma_0((1, Z)) - \gamma_0((0, Z)) \\
    &\ \ \ \ \ \ \ \ \ \ \ \ \ \ \ \ \ \ - \p{\frac{D}{\pi(Z)} - \frac{1 - D}{1 - \pi(Z)}}\Bigp{Y - \gamma(X)} - \gamma((1, Z)) + \gamma((0, Z))}^2}\\
    &= \BiggExp{\Biggp{\p{\frac{D}{\pi(Z)} - \frac{1 - D}{1 - \pi(Z)}}\Bigp{\gamma((D, Z)) - \gamma_0((D, Z))}\\
    &\ \ \ \ \ \ \ \ \ \ \ \ \ \ \ \ \ \ + \gamma_0((1, Z)) - \gamma_0((0, Z)) - \gamma((1, Z)) + \gamma((0, Z))}^2}\\
    &= \BiggExp{\Biggp{\p{1 - \frac{D}{\pi(Z)}}\Bigp{\gamma_0((1, Z)) - \gamma((1, Z))} - \p{1 - \frac{1 - D}{1 - \pi(Z)}}\Bigp{\gamma_0((0, Z)) - \gamma((0, Z))}}^2}.
\end{align*}
Therefore, we have
\begin{align*}
    &\BigExp{\Bigp{h^{\mathrm{AIPW}}(W; \eta_0) - h^{\mathrm{AIPW}}(W; \eta)}^2}\\
    &= \Exp{\p{\p{\frac{D}{\pi_0(Z)} - \frac{1 - D}{1 - \pi_0(Z)}} - \p{\frac{D}{\pi(Z)} - \frac{1 - D}{1 - \pi(Z)}}}^2\Bigp{Y - \gamma_0(X)}^2}\\
    &\ \ \ \ \ + \BiggExp{\Biggp{\p{1 - \frac{D}{\pi(Z)}}\Bigp{\gamma_0((1, Z)) - \gamma((1, Z))} - \p{1 - \frac{1 - D}{1 - \pi(Z)}}\Bigp{\gamma_0((0, Z)) - \gamma((0, Z))}}^2}.
\end{align*}

\section{Direct Debiased Machine Learning}
\label{sec:directdebiasedmachinelearning}
By generalizing the procedure in Section~\ref{sec:basic_idea}, we formulate a direct debiased machine learning framework. 

\subsection{Problem Formulation}
We consider data that consist of observations $\{W_i\}_{i=1}^n$, where each $W_i$ is an i.i.d. copy of $W$ following a distribution $F_0$. The variable $W$ includes an outcome variable $Y$ and regressors $X$; that is $W = (X, Y)$. We focus on parameters that depend on the regression function (conditional mean) of $Y$ given $X$. We denote the true regression function under $F_0$ by $\gamma_0(x) \coloneqq \Exp{Y\mid X = x}$, while we denote a model of $\gamma_0$ by $\gamma$. 

We aim to estimate a parameter of interest of the form
\[\theta_0 \coloneqq \bigExp{m(W, \gamma_0)},\]
where $m(w, \gamma)$ is a functional that depends on a data observation $w$ and a candidate regression function $\gamma$. 

For simplicity, we assume that the expected functional $\gamma \mapsto \Exp{m(W, \gamma)}$ is linear and continuous in $\gamma$, which implies that there is a constant $C > 0$ such that $\Exp{m(W, \gamma)}^2 \leq C \Exp{\gamma(X)^2}$ holds for all $\gamma$ with $\Exp{\gamma(X)^2} < \infty$. Then, from the Riesz representation theorem, there exists a function $v_m$ with $\Exp{v_m(X)^2} < \infty$ such that
\[\Exp{m(W, \gamma)} = \Exp{v_m(X)\gamma(X)}\]
for all $\gamma$ with $\Exp{\gamma(X)^2} < \infty$. This formulation follows \citet{Chernozhukov2022automaticdebiased} and can be generalized to non-linear maps $\gamma \mapsto \Exp{m(W, \gamma)}$. The function $v_m$ is often referred to as the Riesz representer. 

\subsection{Neyman Orthogonal Scores}
Let $\eta_0=(\gamma_0,\alpha_0)$, where $\alpha_0$ is the Riesz representer associated with $m$. The \emph{Neyman orthogonal score} (often equal to the semiparametric efficient influence function) is
\[
\psi(W;\eta,\theta) \coloneqq m(W,\gamma) + \alpha(X)\{Y-\gamma(X)\} - \theta.
\]
It obeys $\Exp{\psi(W;\eta_0,\theta_0)}=0$ and the Neyman orthogonality (Gateaux derivative) with respect to $\eta$ at $\eta_0$ vanishes:
\[
\partial_{\eta}\Exp{\psi(W;\eta,\theta_0)}\big|_{\eta=\eta_0}=0.
\]
Orthogonality ensures that first-order errors from estimating $\eta_0$ do not inflate the asymptotic distribution of the final $\theta$-estimator, provided cross-fitting and mild rate conditions hold; see \citet{Chernozhukov2018doubledebiased,Chernozhukov2022automaticdebiased}. 

\subsection{Direct Debiased Machine Learning}
We propose DDML, which estimates $\eta_0=(\gamma_0,\alpha_0)$ \emph{end-to-end} by directly targeting the oracle score $\psi(W;\eta_0,\theta_0)$. DDML has two pillars:
\begin{enumerate}
    \item \textbf{Neyman targeted estimation}: estimate $(\gamma_0,\alpha_0)$ by minimizing a discrepancy between the oracle score and its plug-in counterpart.
    \item \textbf{Generalized Riesz regression}: estimate $\alpha_0$ via Bregman divergence minimization, which strictly generalizes squared-loss Riesz regression and covers tailored-loss covariate balancing and DRE.
\end{enumerate}

We obtain $\widehat{\theta}$ as the value satisfying 
\[\frac{1}{n}\sum^n_{i=1}\psi(W_i;\widehat{\eta},\widehat{\theta})=0,\]
where $\widehat{\eta}$ denotes estimates of $\eta_0$ plugged into the Neyman orthogonal estimating equations. 

\subsection{Theoretical Properties}
We can show minimax-optimal convergence rates for the nuisance parameter estimators. For details, see \citet{Kato2025directbias} and related studies such as \citet{Chernozhukov2024automaticdebiased}, \citet{Kanamori2012statisticalanalysis}, and \citet{Kato2021nonnegativebregman}.

If the estimators satisfy suitable convergence rate conditions and either the Donsker condition holds or cross-fitting is employed, the resulting estimator $\widehat{\theta}$ is asymptotically normal with asymptotic variance equal to the efficiency bound.

\subsection{Model Specification and Automatic Covariate Balancing}
Under specific choices of Riesz representer models, we can attain covariate balancing without explicitly solving a covariate balancing objective. For example, in ATE estimation, consider the logistic model for the propensity score
\[\pi_\beta(Z) \coloneqq \frac{1}{1 + \exp\bigp{\beta^\top \Phi(Z)}},\]
and define the Riesz representer model as 
\[\alpha_\beta(X) = \frac{\mathbbm{1}[D_i = 1]}{\pi_\beta(X)} - \frac{\mathbbm{1}[D_i = 0]}{1 - \pi_\beta(X)}.\]
In this case, if we use the KL divergence, we obtain the following covariate balancing:
\[\frac{1}{n}\sum^n_{i=1}\alpha_\beta(X_i) \Phi(Z_i) = \frac{1}{n}\sum^n_{i=1}\p{\frac{\mathbbm{1}[D_i = 1]}{\pi_{\widehat{\beta}}(Z_i)} \Phi(Z_i) - \frac{\mathbbm{1}[D_i = 0]}{1 - \pi_{\widehat{\beta}}(Z_i)} \Phi(Z_i)} = 0.\]
We call this property automatic covariate balancing.

\section{Neyman Targeted Estimation}
\label{sec:neymantargetedestimation}
Given any candidates $\eta = (\gamma,\alpha)$ for $\eta_0$ and $\theta^*$ for $\theta_0$, we have
\begin{align*}
&\psi(W;\eta_0,\theta_0)-\psi(W;\eta,\theta^*)\\ 
&= \underbrace{\big(\alpha_0(X)-\alpha(X)\big)\{Y-\gamma_0(X)\}}_{\text{Riesz error}}\quad - \underbrace{\alpha(X)\big(\gamma_0(X)-\gamma(X)\big) - m(W, \gamma) + \theta^*}_{\text{regression error}} + \underbrace{m(W,\gamma_0)-\theta_0}_{\text{mean-zero}}.
\end{align*}
This is derived from the following computation:
\begin{align*}
    &\psi(W, \eta_0, \theta_0) - \psi(W, \eta, \theta^*)\\
    &= m(W, \gamma_0) + \alpha_0(X)\bigp{Y - \gamma_0(X)} - \theta_0 - m(W, \gamma) - \alpha(X)\bigp{Y - \gamma(X)} + \theta^*\\
    &= m(W, \gamma_0) + \alpha_0(X)\bigp{Y - \gamma_0(X)} - \theta_0 - m(W, \gamma)\\
    &\ \ \ \ \ \ \ - \alpha(X)\bigp{Y - \gamma_0(X)} + \alpha(X)\bigp{Y - \gamma_0(X)} - \alpha(X)\bigp{Y - \gamma(X)} + \theta^*\\
    &= \bigp{\alpha_0(X) - \alpha(X)} \bigp{Y - \gamma_0(X)} - \alpha(X)\bigp{\gamma_0(X) - \gamma(X)}\\
    &\ \ \ \ \ \ \ - m(W, \gamma) + \theta^* + m(W, \gamma_0) - \theta_0.
\end{align*}

Thus, targeting $\eta$ amounts to reducing two interpretable components: (i) estimation error for $\alpha_0$, and (ii) estimation error for $\gamma$. Our algorithm alternates or jointly optimizes a Bregman objective (for $\alpha$) and a weighted-risk objective (for $\gamma$).

\begin{itemize}
    \item Estimation of the Riesz representer by Generalized Riesz regression.
    \begin{itemize}
        \item Desirable property: covariate balancing. 
    \end{itemize}
    \item Estimation of the regression functions.
    \begin{itemize}
        \item Two approaches: double estimation and targeted maximum likelihood estimation.
    \end{itemize}
\end{itemize}

\subsection{Estimation of the Riesz Representer}
We estimate the Riesz representer by minimizing the error 
\[\abs{\bigp{\alpha_0(X) - \alpha(X)} \bigp{Y - \gamma_0(X)} - \alpha(X)\bigp{\gamma_0(X) - \gamma(X)}}.\]
Since $\alpha(X)\bigp{\gamma_0(X) - \gamma(X)} = 0$ if $\gamma = \gamma_0$, we focus on $\bigp{\alpha_0(X) - \alpha(X)} \bigp{Y - \gamma_0(X)}$. 

\paragraph{Generalized Riesz Regression.} There exist several measures for evaluating $\bigp{\alpha_0(X) - \alpha(X)} \bigp{Y - \gamma_0(X)}$. In this study, we propose using the Bregman divergence to measure the discrepancy between $\alpha_0(X)$ and $\alpha(X)$, weighted by $\Exp{Y - \gamma_0(X)\mid X}^2$, the conditional variance of $Y$ given $X$. We explain the details of the Riesz representer estimation via the Bregman divergence in Section~\ref{sec:grr}. We refer to this estimation method as Generalized Riesz regression. Here, we briefly introduce the idea of Bregman divergence estimation. 

The Bregman divergence depends on a convex function, and by specifying this convex function, we obtain various divergence metrics, including loss functions such as the squared loss and log loss. For example, under the squared loss at the population level, we estimate the Riesz representer as 
\[\alpha^* \coloneqq \argmin_{\alpha \in \calA}\Exp{\bigp{\alpha_0(X) - \alpha(X)}^2 \bigp{Y - \gamma_0(X)}^2},\]
where $\calA$ is a set of candidates of $\alpha_0$. If $\alpha_0 \in \calA$, then $\alpha^* = \alpha_0$ holds. 
This objective includes the unknown $\alpha_0(X)$ (we replace $\gamma_0$ with a consistent estimator, as it only weights the loss). Although the presence of $\alpha_0(X)$ seems to make the optimization infeasible, we can obtain an equivalent objective that does not include $\alpha_0(X)$:
\[\alpha^* \coloneqq \argmin_{\alpha \in \calA}\BigExp{ - g(\alpha(X)) + \partial g(\alpha(X)) \alpha(X) - m\bigp{\partial g(\alpha(X))}},\]
where $g$ differentiable and strictly convex function.
Estimation of $\alpha_0$ based on this optimization is referred to as Riesz regression in automatic debiased machine learning \citep{Chernozhukov2024automaticdebiased}; when $\alpha_0(X)$ is related to a density ratio, it is a special case of LSIF in the density-ratio estimation literature \citep{Kanamori2009aleastsquares}. 

The Bregman divergence also includes other loss functions. In particular, under specific choices, we obtain the covariate balancing propensity score \citep{Imai2013estimatingtreatment} and empirical balancing \citep{Hainmueller2012entropybalancing} by bypassing the tailored loss argument of \citet{Zhao2019covariatebalancing}, which also shows the equivalence between covariate balancing propensity scores and empirical balancing.

\paragraph{Automatic Covariate Balancing.}
By selecting different convex functions in the Bregman divergence, Generalized Riesz regression induces different losses. In treatment effect estimation, certain choices yield a desirable property we call automatic covariate balancing. We provide details in Section~\ref{sec:covariatebalancingconsistency}. 

\subsection{Estimation of the Regression Function}
There are two main approaches to estimating the regression functions. 

\paragraph{Case~1: Double Estimation} Estimate the regression function $\gamma_0$ and the Riesz representer separately, using the two-step or iterative procedures in Section~\ref{sec:basic_idea}. 

\paragraph{Case~2: TMLE}
After double estimation, given estimates $\widehat{\gamma}^{(0)}$ and $\widehat{\alpha}$, update the regression function as
\[\widehat{\gamma}^{(1)} \coloneqq \widehat{\gamma}^{(0)} + \frac{\sum_{i=1}^n \widehat{\alpha}(X_i)\bigp{Y_i - \widehat{\gamma}(X_i)}}{\sum_{i=1}^n \widehat{\alpha}(X_i)^2}\widehat{\alpha}(X_i).\]
Then estimate the parameter of interest as 
\[\widehat{\theta}^{\text{TMLE}} \coloneqq \frac{1}{n}\sum^n_{i=1}\p{\widehat{\gamma}^{(1)}((1, Z_i)) - \widehat{\gamma}^{(1)}((0, Z_i))} = \frac{1}{n}\sum^n_{i=1}m(X_i, \widehat{\gamma}^{(1)}(X_i)).\]
Note that this update is derived as the solution in $\epsilon$ of
\[\sum^n_{i=1}\p{Y_i - \p{\widehat{\gamma}^{(0)}(X_i) + \epsilon\widehat{\alpha}(X_i)}} = 0,\]
which is given as $\widehat{\epsilon} \coloneqq 
\frac{\sum_{i=1}^n \widehat{\alpha}(X_i)\{Y_i-\widehat{\gamma}^{(0)}(X_i)\}}
     {\sum_{i=1}^n \widehat{\alpha}(X_i)^2},\qquad
\widehat{\gamma}^{(1)}(x)=\widehat{\gamma}^{(0)}(x)+\widehat{\epsilon}\widehat{\alpha}(x)$. 

Under this update, for $\widehat{\eta} = (\widehat{\gamma}, \widehat{\pi})$ and $\widehat{\theta}^{\text{TMLE}} = m(X_i, \widehat{\gamma}^{(1)})$, we have 
\begin{align*}
    &\psi(W_i, \eta_0, \theta_0) - \psi(W_i, \widehat{\eta}, \widehat{\theta}^{\text{TMLE}} )\\
    &= \bigp{\alpha_0(X_i) - \widehat{\alpha}(X_i)} \bigp{Y_i - \gamma_0(X_i)} - \alpha(X_i)\bigp{\gamma_0(X_i) - \widehat{\gamma}^{(1)}(X_i)}\\
    &\ \ \ \ \ - m(X_i, \widehat{\gamma}^{(1)}) + \widehat{\theta}^{\text{TMLE}} + m(X_i, \gamma_0) - \theta_0\\
    &= \bigp{\alpha_0(X_i) - \widehat{\alpha}(X_i)} \bigp{Y_i - \gamma_0(X_i)} - \alpha(X_i)\bigp{Y_i - \widehat{\gamma}^{(1)}(X_i)}\\
    &\ \ \ \ \ \annot{- m(X_i, \widehat{\gamma}^{(1)}) + \widehat{\theta}^{\text{TMLE}}}{$=0$ in sample mean} + \annot{m(X_i, \gamma_0) - \theta_0 + \alpha(X_i)\bigp{Y_i - \gamma_0(X_i)}}{$=0$ in expectation}.
\end{align*}
Therefore, 
\begin{align*}
    &\Exp{\frac{1}{n}\sum^n_{i=1}\p{\psi(W_i, \eta_0, \theta_0) - \psi(W, \eta, \theta)}}\\
    &= \Exp{\frac{1}{n}\sum^n_{i=1}\p{\bigp{\alpha_0(X_i) - \widehat{\alpha}(X_i)} \bigp{Y_i - \gamma_0(X_i)}}}.
\end{align*}
Thus, the estimation error between the oracle and plug-in Neyman orthogonal scores boils down to the estimation error of the Riesz representer. Hence, Neyman targeted estimation reduces to estimating the Riesz representer.
\section{Generalized Riesz Regression}
\label{sec:grr}
The essential idea in Section~\ref{sec:basic_idea} is to estimate $\gamma_0$ and $\pi_0$ by minimizing the error between $h^{\mathrm{AIPW}}(W; (\gamma_0, \pi_0)) - h^{\mathrm{AIPW}}(W; (\gamma, \pi))$. While we measure the error using the squared loss in Section~\ref{sec:basic_idea}, here we measure the error using the Bregman divergence. Bregman divergence is a general measure of deviation between two quantities. It not only includes the MSE but also includes the KL divergence as a special case, where the KL divergence can be interpreted as a likelihood. We then formulate the nuisance parameter estimation problem as a Bregman divergence minimization problem. 

Through the lens of Bregman divergence minimization, we find several useful connections. As noted above, when using the squared loss, the target estimation method includes Riesz regression in \citet{Chernozhukov2024automaticdebiased} as a special case. In addition, when using the KL divergence, we can interpret the method as the tailored loss in covariate balancing \citep{Zhao2019covariatebalancing}, which includes the covariate balancing propensity scores and entropy balancing as special cases. Furthermore, techniques used in our arguments are based on the density-ratio estimation literature. For example, Riesz regression is essentially the same formulation as LSIF in \citet{Kanamori2009aleastsquares}, and the generalization by Bregman divergence appears in \citet{Sugiyama2011densityratio}. Thus, Riesz regression, covariate balancing, and density-ratio estimation essentially share the same formulation. 

\subsection{Bregman Divergence}
Let $g\colon \bbR \to \bbR$ be a differentiable and strictly convex function. Given $x \in \calX$, define the Bregman divergence between $\alpha_0(x), \alpha(x) \colon \calX \to \bbR$ as
\[\text{BR}^\dagger_g\bigp{\alpha_0(x)\mid \alpha(x)} \coloneqq g(\alpha_0(x)) - g(\alpha(x)) - \partial g(\alpha(x)) \bigp{\alpha_0(x) - \alpha(x)},\] 
where $\partial g$ denotes the derivative of $g$. Then define the average Bregman divergence as
\[\text{BR}^\dagger_g\bigp{\alpha_0\mid \alpha} \coloneqq \BigExp{g(\alpha_0(X)) - g(\alpha(X)) - \partial g(\alpha(X)) \bigp{\alpha_0(X) - \alpha(X)}}.\] 
We estimate $\alpha_0$ by 
\[\alpha^* = \argmin_{\alpha\in \calA} \text{BR}^\dagger_g\bigp{\alpha_0\mid \alpha}.\]
where $\calA$ is a set of candidates of $\alpha_0$. If $\alpha_0 \in \calA$, then $\alpha^* = \alpha_0$ holds.

Although $\alpha_0$ is unknown, we can define an equivalent optimization problem without using $\alpha_0$:
\[\alpha^* = \argmin_{r\in \calA} \text{BR}_g\bigp{\alpha},\] 
where 
\[\text{BR}_g\bigp{\alpha} \coloneqq \BigExp{ - g(\alpha(X)) + \partial g(\alpha(X)) \alpha(X) - m\bigp{\partial g(\alpha(X))}}.\]
Here, we used the linearity of $m$
\[
\E\big[\partial g(\alpha(Z))\alpha_0(Z)\big]=\E\big[m\{\partial g(\alpha(Z))\}\big].
\]
By choosing different $g$, we derive various objectives for Riesz representer estimation, including Riesz regression.

\subsection{Empirical Bregman Divergence Minimization}
We estimate the Riesz representer $\alpha_0$ by minimizing an empirical Bregman divergence:
\[
\widehat{\alpha}_n \coloneqq \argmin_{\alpha \in \calH}\widehat{\text{BR}}_g\bigp{\alpha}  + \lambda J(\alpha),
\]
where $J(\alpha)$ is some regularization function, and
\[
\widehat{\text{BR}}_g(\alpha) \coloneqq \frac{1}{n}\sum^n_{i=1}\Bigp{ - g(\alpha(X_i)) + \partial g(\alpha(X_i)) \alpha(X_i) - m\bigp{\partial g(\alpha(X_i))}}.
\]

\subsection{Weighted Empirical Bregman Divergence Minimization}
We can also define a weighted version of the Bregman divergence minimization:
\[
\widehat{\alpha}_n \coloneqq \argmin_{\alpha \in \calH}\widehat{\text{BR}}^W_g\bigp{\alpha}  + \lambda J(\alpha),
\]
where
\[
\widehat{\text{BR}}^W_g(\alpha) \coloneqq \frac{1}{n}\sum^n_{i=1}\Bigp{ - g(\alpha(X_i)) + \partial g(\alpha(X_i)) \alpha(X_i) - m\bigp{\partial g(\alpha(X_i))}}\Bigp{Y_i - \widehat{\gamma}(X_i)}^2,
\]
and $\widehat{\gamma}$ is some estimate of $\gamma_0$. This weighted version multiplies each summand by $\{Y-\widehat{\gamma}(D,Z)\}^2$ to stabilize variance and improve small-sample behavior.

\section{ATE Estimation}
\label{sec:ateest}
We revisit the ATE estimation. The parameter of interest is the ATE, defined as $\tau_0 = \Exp{m^{\text{ATE}}(W, \gamma_0)}$, where
\[m^{\text{ATE}}(W, \gamma_0) \coloneqq \gamma_0((1, Z)) - \gamma_0((0, Z)),\]
where $\gamma_0((d, Z)) = \Exp{Y\mid D = d, Z}$. 
In this setting, the Riesz representer is
\[\alpha^{\text{ATE}}_0(X) \coloneqq \frac{D}{\pi_0(Z)} - \frac{1 - D}{1 - \pi_0(Z)},\]
where $\pi_0(Z) = \Pr(D = 1\mid Z)$ is the propensity score. Then the Neyman orthogonal score is
\[\psi(W; \eta_0, \tau_0) = h^{\text{AIPW}}(W; \eta_0) - \tau_0,\]
where recall that $h^{\mathrm{AIPW}}(W; \eta) = \p{\frac{D}{\pi(Z)} - \frac{1 - D}{1 - \pi(Z)}}\Bigp{Y - \gamma(X)} + \gamma((1, Z)) - \gamma((0, Z))$. Then, the estimation error in the Neyman targeted step is given as
\begin{align*}
    &\frac{1}{n}\sum^n_{i=1}\Bigp{h^{\text{AIPW}}(W_i; \eta_0) - h^{\text{AIPW}}(W_i; \eta)}\\
    &= \frac{1}{n}\sum^n_{i=1}\Bigp{h^{\text{AIPW}}(W_i; \eta_0) - h^{\text{AIPW}}(W_i; (\pi, \gamma_0))} + \frac{1}{n}\sum^n_{i=1}\Bigp{h^{\text{AIPW}}(W_i; (\pi, \gamma_0)) - h^{\text{AIPW}}(W_i; (\pi, \gamma))}.
\end{align*}

\subsection{Least Squares}
We first introduce the squared function for the Bregman divergence, given as
\[g^{\text{LS}}(\alpha) = (\alpha - 1)^2.\]
Under this choice of $g$, we have
\[\text{BR}_{g^{\text{LS}}}\bigp{\alpha} = \Exp{- 2\bigp{\alpha((1, Z)) + \alpha((0, Z))} + \mathbbm{1}[D = 1]\alpha((1, Z))^2 + \mathbbm{1}[D = 0]\alpha((0, Z))^2 }.\]
Then, we estimate $\alpha_0$ by minimizing the empirical objective:
\[
\widehat{\alpha} \coloneqq \argmin_{\alpha \in \calA}\widehat{\text{BR}}_{g^{\mathrm{LS}}}\bigp{\alpha},
\]
where 
\[\widehat{\text{BR}}_{g^{\mathrm{LS}}}\bigp{\alpha} \coloneqq \frac{1}{n}\sum^n_{i=1}\p{- 2\bigp{\alpha((1, Z_i)) + \alpha((0, Z_i))} + \mathbbm{1}[D_i = 1]\alpha((1, Z_i))^2 + \mathbbm{1}[D_i = 0]\alpha((0, Z_i))^2 }.\]
This objective function matches LSIF for density-ratio estimation proposed in \citet{Kanamori2009aleastsquares,Kanamori2012statisticalanalysis} and the Riesz regression for automatic debiased machine learning proposed in \citet{Chernozhukov2024automaticdebiased}. Furthermore, under specific choice of $\calA$, a model of $\alpha_0$, this objective function also yields nearest neighbor matching, as \citet{Kato2025nearestneighbor} shows the equivalence between LSIF (equivalently, Riesz regression) and density-ratio estimation method proposed in \citet{Lin2023estimationbased}. \citet{Kato2022learningcausal} also applies LSIF for nonparametric instrumental variable regression.

For $\alpha$, we can use various models, such as random forests and neural networks. Note that there can be occurred train-loss hacking problems and appropriate correction methods are also required \citep{Rhodes2020telescopingdensityratio,Kato2021nonnegativebregman,Kiryo2017positiveunlabeledlearning}. 

\subsection{Maximum Likelihood Estimation}
\label{sec:constmle}
We can also derive (constrained) maximum likelihood estimation. Let us consider the convex function
\[g^{\mathrm{KL}}(\alpha) = |\alpha|\log |\alpha| - |\alpha|.\]
This choice connects the Bregman divergence to the KL divergence. 
Here, we have
\[\partial g^{\mathrm{KL}}(\alpha(X)) = \begin{cases}
    \log \p{\alpha((1, Z))} & \text{if}\ \ D = 1\\
    -\log \p{- \alpha((0, Z))}& \text{if}\ \ D = 0
\end{cases}.\]

Substituting this $g^{\mathrm{KL}}$ into the Bregman divergence gives
\[\text{BR}_{g^{\mathrm{KL}}}\bigp{\alpha} \coloneqq \bigExp{ \sign(\alpha(X))\log|\alpha(X)| - \log \alpha((1, Z)) - \log \alpha((0, Z))}.\]
Then, we estimate $\alpha_0$ by minimizing the empirical objective:
\[
\widehat{\alpha} \coloneqq \argmin_{\alpha \in \calA}\widehat{\text{BR}}_{g^{\mathrm{KL}}}\bigp{\alpha}  + \lambda J(\alpha),
\]
where 
\[\widehat{\text{BR}}_{g^{\mathrm{KL}}}\bigp{\alpha} \coloneqq \frac{1}{n}\sum^n_{i=1}\p{ \big|\alpha(X_i)\big| - \log \big|\alpha((1, Z_i))\big| - \log \big|\alpha((0, Z_i))\big|}.\]
This corresponds to unnormalized KL (UKL) divergence minimization in density-ratio estimation \citep{Sugiyama2011densityratio}, which generalizes Kullback-Leibler Importance Estimation. Procedure \citep[KLIEP,][]{Sugiyama2008directimportance,Nguyen2007estimatingdivergence}. 

\paragraph{Inverse propensity score modeling}
Consider modeling the Riesz representer via a model of the inverse propensity score. Let $r((d, Z))$ model the inverse propensity score, that is, $r((1, Z))$ models $r_0((1, Z)) \coloneqq 1/\pi_0(Z)$ and $r((0, Z))$ models $r_0((0, Z)) \coloneqq 1/(1 - \pi_0(Z))$. Using this model, define
\[\alpha_r((D, Z)) = \mathbbm{1}[D = 1]r((1, Z)) - \mathbbm{1}[D = 0]r((0, Z)).\]

By definition, we use $r$ such that $r(X) \in (1, \infty)$. Then, we have $\alpha((1, Z)) = r((1, Z)) \in (1, \infty)$ and $\alpha((0, Z)) = -r((0, Z)) \in (-\infty, -1)$. Therefore, we have 
\begin{align*}
    g^{\mathrm{KL}}(\alpha(X)) = \begin{cases}
        r((1, Z))\log r((1, Z)) - r((1, Z)) & \text{if}\ D = 1\\
        r((0, Z))\log r((0, Z)) - r((0, Z)) & \text{if}\ D = 0
    \end{cases}.
\end{align*}

Then we obtain the objective
\[\text{BR}_{g^{\mathrm{KL}}}\bigp{\alpha_r} \coloneqq \bigExp{ -\log(r((1, Z))) - \log(r((0, Z))) + \mathbbm{1}[D_i = 1]r((1, Z_i)) + \mathbbm{1}[D_i = 0]r((0, Z_i))}.\]
Let $\calR$ be a set of $r$. Then, we estimate 
\begin{align}
    \label{eq:mle}
    r^* \coloneqq \argmin_{r \in \calR}\text{BR}_{g^{\mathrm{KL}}}\bigp{\alpha_r}
\end{align}
Since we do not know the expected value, by replacing it with an empirical estimate, we estimate $r_0$ in the Riesz representer by 
\begin{align*}
    \widehat{r}_n \coloneqq \argmin_{r \in \calR}\widehat{\text{BR}}_{g^{\mathrm{KL}}}\bigp{\alpha_r} + \lambda J(\alpha),
\end{align*}
where 
\begin{align*}
    &\widehat{\text{BR}}_{g^{\mathrm{KL}}}(\alpha_r) \coloneqq\\
    &\frac{1}{n}\sum^n_{i=1}\bigp{ -\log \bigp{r((1, Z_i))} - \log \bigp{r((0, Z_i))} + \mathbbm{1}[D_i = 1]r((1, Z_i)) + \mathbbm{1}[D_i = 0]r((0, Z_i))}
\end{align*}

Solving \eqref{eq:mle} is equivalent to 
\begin{align*}
    r^* &= \argmax_{r\in\calR} \bbE\sqb{\log r((1, Z)) + \log r((0, Z))}\\
    &\mathrm{s.t.}\quad \bigExp{\mathbbm{1}[D = 1]r((1, Z))} = \bigExp{\mathbbm{1}[D = 0]r((0, Z))} = 1.
\end{align*}
This technique is known as Silverman's trick \citep{Silverman1982onestimation}. For details, see Theorem~3.3 in \citet{Kato2023unifiedperspective}. Replacing expectations with sample means yields the estimation problem
\begin{align*}
    \widehat{r}_n &= \argmax_{r\in\calR} \frac{1}{n}\sum^n_{i=1}\bigp{\log r((1, Z_i)) + \log r((0, Z_i))}\\
    &\mathrm{s.t.}\quad \frac{1}{n}\sum^n_{i=1}\mathbbm{1}[D_i = 1]r((1, Z_i)) = \frac{1}{n}\sum^n_{i=1}\mathbbm{1}[D_i = 0]r((0, Z_i)) = 1
\end{align*}

\subsection{Empirical Balancing}
\label{sec:empbalancing}
Next, we derive empirical balancing as a special case of Bregman divergence minimization. Empirical balancing is a specific form of covariate balancing and can be derived from a tailored loss function \citep{Zhao2019covariatebalancing}.

Consider a convex function defined as
\[g^{\mathrm{E}}(\alpha) = (|\alpha|-1)\log\p{\abs{\alpha} - 1} - |\alpha|.\]
This choice also yields another KL-type divergence.
Note that $\alpha < 0$ or $\alpha > 1$ always holds, and for $\alpha \in (-\infty, 0)$ and $\alpha \in (1, \infty)$, $g^{\mathrm{E}}$ is convex with derivative
\[\partial g^{\mathrm{E}}(\alpha) = \begin{cases}
    - \log\p{-\alpha - 1} & \text{if}\ \alpha < 0\\
    \log\p{\alpha - 1} & \text{if}\ \alpha > 1
\end{cases}.\]
Substituting this $g^{\mathrm{E}}$ and $\alpha(X) = \mathbbm{1}[D = 1]r((1, Z)) + \mathbbm{1}[D = 0]r((0, Z))$, we obtain
\begin{align*}
    &\text{BR}_{g^{\mathrm{E}}}\bigp{\alpha} \coloneqq \BigExp{\log\bigp{|\alpha(X)| - 1} + |\alpha(X)| - \log\bigp{\alpha((1, Z)) - 1} - \log\bigp{-\alpha((0, Z)) - 1}}.
\end{align*}
Then, we estimate $\alpha_0$ by minimizing the empirical objective:
\[
\widehat{\alpha} \coloneqq \argmin_{\alpha \in \calA}\widehat{\text{BR}}_{g^{\mathrm{E}}}\bigp{\alpha}  + \lambda J(\alpha),
\]
where 
\[\widehat{\text{BR}}_{g^{\mathrm{E}}}\bigp{\alpha} \coloneqq \frac{1}{n}\sum^n_{i=1}\Bigp{\log\p{|\alpha(X_i)| - 1} + |\alpha(X_i)| - \log\p{\alpha((1, Z_i)) - 1} - \log\p{-\alpha((0, Z_i)) - 1}}.\]

\paragraph{Inverse propensity score modeling}
As in Section~\ref{sec:constmle}, model the Riesz representer via the inverse propensity score. Let $r((d, Z))$ model the inverse propensity score, that is, $r((1, Z))$ models $r_0((1, Z)) \coloneqq 1/\pi_0(Z)$ and $r((0, Z))$ models $r_0((0, Z)) \coloneqq 1/(1 - \pi_0(Z))$. Based on this model, define 
\[\alpha_r(X) = \mathbbm{1}[D_i = 1]r((1, Z)) - \mathbbm{1}[D_i = 0]r((0, Z)).\]
Under this model, we write the Bregman divergence as
\[\text{BR}_{g^{\mathrm{E}}}\bigp{\alpha_r} \coloneqq \sum_{d \in \{1, 0\}}\BigExp{\mathbbm{1}[D = d]\Bigp{\log\p{r((d, Z)) - 1} + r((d, Z))} - \log\p{r((d, Z)) - 1}}.\]
We can simplify this Bregman divergence as
\begin{align*}
&\text{BR}_{g^{\mathrm{E}}}\bigp{\alpha_r} = \BigExp{- \mathbbm{1}[D = 0]\log\p{r((1, Z))-1} - \mathbbm{1}[D = 1]\log\p{r((0, Z))-1}\\
&\qquad\qquad\qquad\qquad\qquad\qquad\qquad\quad + \mathbbm{1}[D = 1]r((1, Z)) + \mathbbm{1}[D = 0]r((0, Z))}.
\end{align*}
Then, we estimate $r_0$ in the Riesz representer by
\[
\widehat{r}_n \coloneqq \argmin_{r \in \calR}\widehat{\text{BR}}_{g^{\mathrm{E}}}\bigp{\alpha_r},
\]
where the empirical Bregman divergence is
\begin{align*}
    &\widehat{\text{BR}}_{g^{\mathrm{E}}}(\alpha_r) \coloneqq \frac{1}{n}\sum^n_{i=1}\bigp{\mathbbm{1}[D_i = 0]\log \bigp{r((1, Z_i)) - 1} + \mathbbm{1}[D_i = 1]\log \bigp{r((0, Z_i)) - 1}\\
    &\qquad\qquad\qquad\qquad\qquad\qquad\qquad + \mathbbm{1}[D_i = 1]r((1, Z_i)) + \mathbbm{1}[D_i = 0]r((0, Z_i))}.
\end{align*}

\paragraph{Connection to the tailored loss.}
Since $r((1, Z)) = 1/\pi(X)$ and $r((0, Z)) = 1/(1 - \pi(X))$, we have $r((1, Z)) - 1 = 1/\big(r((0, Z)) - 1\big)$. Therefore, it holds that
\begin{align*}
    &\widehat{\text{BR}}_{g^{\mathrm{E}}}(\alpha_r) = \frac{1}{n}\sum^n_{i=1}\Biggp{\mathbbm{1}[D_i = 1]\p{ - \log \p{\frac{1}{r((1, Z_i)) - 1}} + r((1, Z_i))}\\
    &\qquad\qquad\qquad\qquad\qquad\qquad\qquad + \mathbbm{1}[D_i = 0]\p{ - \log \p{\frac{1}{r((0, Z_i)) - 1}} + r((0, Z_i))}}.
\end{align*}
This objective is equivalent to the tailored loss in \citet{Zhao2019covariatebalancing}. From this objective, we can also derive empirical balancing \citep{Chan2015globallyefficient}.

\section{ATT Estimation}
\label{sec:att}
We consider the ATT,
\[
\tau^{\text{ATT}}_0 \coloneqq \Exp{Y(1)-Y(0)\mid D=1}
= \frac{1}{p}\Exp{ \gamma_0((1,Z))-\gamma_0((0,Z)) \mid D=1},
\]
where $p\coloneqq \Pr(D=1)$ and, as in Section~\ref{sec:basic_idea}, $\gamma_0((d,Z))=\Exp{Y\mid (D, Z) = (d, Z)}$.
Define the linear functional
\[
m^{\text{ATT}}(W,\gamma)\coloneqq \frac{D}{p}\Bigp{\gamma((1,Z))-\gamma((0,Z))},\qquad
\tau^{\text{ATT}}_0 = \Exp{m^{\text{ATT}}(W,\gamma_0)}.
\]
Its Riesz representer $\alpha^{\text{ATT}}_0$ is characterized by $\bigExp{m^{\text{ATT}}(W,\gamma)} = \bigExp{\alpha^{\text{ATT}}_0(X)\gamma(X)}$ for all square–integrable $\gamma$. A standard calculation yields
\[
\alpha^{\text{ATT}}_0(X)=\frac{D}{p}-\frac{1-D}{p}\cdot\frac{\pi_0(Z)}{1-\pi_0(Z)},
\]
where $\pi_0(Z)=\Pr(D=1\mid Z)$.

\subsection{Neyman Orthogonal Scores}
The Neyman orthogonal score is
\[
\psi^{\text{ATT}}\p{W;\eta,\tau^{\text{ATT}}}\coloneqq m^{\text{ATT}}(W,\gamma)+\alpha^{\text{ATT}}(X)\Bigp{Y-\gamma(X)}-\tau^{\text{ATT}},
\]
with $\eta=(\gamma,\alpha^{\text{ATT}})$. Plugging in estimates gives a DDML estimator solving
\[
\frac{1}{n}\sum_{i=1}^n\psi^{\text{ATT}}\p{W_i;\widehat\eta,\widehat{\tau}^{\text{ATT}}}=0,
\]
which is root-$n$ and efficient under mild convergence–rate and complexity conditions (via the Donsker condition or cross–fitting), without requiring an explicit model for $\alpha^{\text{ATT}}_0$.

\subsection{Generalized Riesz Regression}
As in Section~\ref{sec:grr}, estimate $\alpha^{\text{ATT}}_0$ by minimizing an empirical Bregman risk
\[
\widehat{\alpha}^{\text{ATT}}\in\argmin_{\alpha\in\calA}\ \frac1n\sum_{i=1}^n
\Big\{-g(\alpha(X_i))+\partial g(\alpha(X_i))\alpha(X_i)-m^{\text{ATT}}\big(W_i,\partial g(\alpha(\cdot))\big)\Big\}
+\lambda J(\alpha),
\]
optionally with the variance–stabilizing weight $\p{Y_i-\widehat\gamma(X_i)}^2$ as in Section~\ref{sec:grr}. Squared loss $g(\alpha)=(\alpha-1)^2$ recovers Riesz regression, while KL-type $g$ induces entropy–style balancing losses.

\paragraph{Squared loss (LSIF/Riesz regression).}
With $g^{\text{LS}}(\alpha)=(\alpha-1)^2$, the empirical objective reduces to Riesz regression, which also coincides with LSIF. Furthermore, the dual matches the stable weights in \citet{Zubizarreta2015stableweights} when linear models are used for $\alpha$.

\paragraph{KL-type loss (entropy balancing).}
With $g^{\text{KL}}(\alpha)=(|\alpha| - 1)\log(|\alpha| - 1)-|\alpha|$, the dual problem enforces moment balance between treated units and reweighted controls (automatic covariate balancing), aligning with entropy–balancing–style weights but learned through the Riesz objective.

\section{AME Estimation}
\label{sec:ame}
Let $X=(D,Z)$ with a (scalar) continuous treatment $D$, and define the AME as
\[
\theta^{\text{AME}}_0 \coloneqq \Exp{\partial_d \gamma_0((D,Z))}.
\]
Here, linear functional is given as 
\[m^{\text{AME}}(W,\gamma)=\partial_d \gamma((D, Z)).\]
The Riesz representer that satisfies $\Exp{m^{\text{AME}}(W,\gamma)}=\Exp{\alpha^{\text{AME}}_0(X)\gamma(X)}$ is the (negative) score of the joint density of $X = (D, Z)$ with respect to $d$:
\[
\alpha^{\text{AME}}_0(X) = - \partial_d \log f_0((D,Z)),
\]
where $f_0(X)$ is the joint probability density of $X$. 

\subsection{Neyman Orthogonal Scores}
The Neyman orthogonal score is given as
\[
\psi^{\text{AME}}(W; \eta, \theta) = m^{\text{AME}}(W, \gamma)+\alpha^{\text{AME}}(X)\bigp{Y-\gamma(X)}-\theta.
\]
Then, we define the Neyman targeted estimation for estimating $\alpha^{\text{AME}}_0$ and $\gamma_0$ with generalized Riesz regression. 

\subsection{Generalized Riesz Regression}
Generalized Riesz regression estimates the AME Riesz representer $\alpha^{\mathrm{AME}}_0$ directly, without explicitly modeling the density (or its score) $\partial_d\log f_{D\mid Z}(D\mid Z)$.
Recall the population Bregman objective for a differentiable, strictly convex $g$:
\[
\mathrm{BR}_g(\alpha)
\coloneqq
\BigExp{-g\big(\alpha(X)\big)+\partial g\big(\alpha(X)\big)\alpha(X)-m\bigp{\partial g(\alpha)}},
\]
where, for AME, the linear functional is $m(\alpha)=\Exp{\partial_d \alpha(X)}$. Hence $m\big(\partial g(\alpha)\big)=\bigExp{\partial_d\{\partial g(\alpha(X))\}}$.

\paragraph{Squared loss.}
Let $g^{\text{LS}}(u)=(u-1)^2$ with $\partial g^{\text{LS}}(u)=2(u-1)$. Then
\[
\mathrm{BR}_{g^{\text{LS}}}(\alpha)
=
\BigExp{
-\big(\alpha(X)-1\big)^2
+2\big(\alpha(X)-1\big)\alpha(X)
-\partial_d\big\{2\big(\alpha(X)-1\big)\big\}}.
\]
Here, we have the equivalent form (up to an additive constant independent of $\alpha$)
\[
\mathrm{BR}_{g^{\text{LS}}}(\alpha)
=
\bigExp{\alpha(X)^2-2\partial_d \alpha(X)}
+\text{const}.
\]
Thus the squared-loss Bregman objective targets $\alpha^{\text{AME}}_0$ in $L_2$. This method corresponds to Riesz regression, shown in Section~2.2 in \citet{Chernozhukov2024automaticdebiased}. 

\paragraph{KL-type divergence.}
For the signed KL-type convex function 
\[g^{\text{KL}}(\alpha)=|\alpha|\log|\alpha|-|\alpha|,\] 
the AME objective is given as
\begin{align*}
    &\mathrm{BR}_{g^{\text{KL}}}(\alpha) \coloneqq\\
    &\BigExp{
-|\alpha(X)|\log|\alpha(X)|
+|\alpha(X)|
+\mathrm{sign}\bigp{\alpha(X)}\log|\alpha(X)|\alpha(X)
-\partial_d\bigp{\mathrm{sign}(\alpha(X))\log|\alpha(X)|}}.
\end{align*}
Minimizing the empirical version $\widehat{\mathrm{BR}}_g$ over a chosen class $\calA$ gives us an estimator of $\alpha_0$.

\section{Covariate Shift Adaptation}
\label{sec:covshift}
Let $X\sim F_0$ denote a source covariate distribution that generates labeled data $(X,Y)$, and let $\widetilde X\sim G_0$ denote a target covariate distribution. 
Let $\{(X_i,Y_i)\}_{i\in \calI_S}$ be i.i.d. from $F_0$ (source) and $\{\widetilde{X}_j\}_{j\in \calI_T}$ i.i.d.\ from $G_0$ (target), independent.

Suppose we train $\gamma_0(x)=\Exp{Y\mid X=x}$ on a source population with covariates $X\sim F_0$, but the target parameter averages $\gamma_0$ over a \emph{shifted} covariate distribution $\widetilde{X}\sim G_0$:
\[
\theta^{\text{CS}}_0 \coloneqq \bigExp{\gamma_0(\widetilde{X})}
= \bigExp{m^{\text{CS}}(\widetilde{X},\gamma_0)},\qquad m^{\text{CS}}(x,\gamma)\coloneqq \gamma(x).
\]
When $G_0$ is absolutely continuous with respect to $F_0$ with density ratio $r_0(x)\coloneqq \frac{dG_0}{dF_0}(x)$, the Riesz representer is
\[
\alpha^{\text{CS}}_0(X)=r_0(X)=\frac{g_0(X)}{f_0(X)}.
\]

\subsection{Neyman Orthogonal Scores}
The orthogonal score
\[
\psi^{\text{CS}}(W;\eta,\theta)=\gamma(X)-\theta+\alpha^{\text{CS}}(X)\{Y-\gamma(X)\}
\]
delivers a debiased estimator by combining source residuals with target averaging, accommodating independent source/target samples via cross–fitting or data fusion.

\subsection{Generalized Riesz Regression via Bregman Matching}
We estimate $\alpha^{\text{CS}}_0$ directly by density–ratio matching under a Bregman divergence. Let $g:\bbR_+\to\bbR$ be differentiable and strictly convex. The population Bregman risk for a ratio model $\alpha$ is
\[
\text{BR}_g(\alpha)
\coloneqq
\bbE_{G_0}\bigsqb{\partial g(\alpha(X))\alpha(X)-g(\alpha(X))}
 - 
\bbE_{F_0}\bigsqb{\partial g(\alpha(X))},
\]
which is equivalent to the Bregman divergence $\text{BR}'_g(r_0\mid r)$ up to a constant independent of $r$; the same decomposition and its empirical version are standard in the density-ratio estimation framework. 

Given target samples $\{\widetilde{X}_j\}$ and source samples $\{X_i\}$, the empirical counterpart is
\[
\widehat{\text{BR}}_g(\alpha)
\coloneqq
\frac{1}{|\calI_T|}\sum_{j\in\calI_T}\big\{\partial g(\alpha(\widetilde{X}_j))\alpha(\widetilde{X}_j)-g(r(\widetilde{X}_j))\big\}
-
\frac{1}{|\calI_S|}\sum_{i\in\calI_S}\partial g\big(\alpha(X_i)\big),
\]
and we set
\[
\widehat{\alpha} \coloneqq 
\arg\min_{\alpha\in\calA}
\widehat{\text{BR}}_g(\alpha)+\lambda J(\alpha),
\]
with a model class $\calR$ and regularizer $J$.

\paragraph{Squared loss.}
For $g^{\text{LS}}(\alpha)=(\alpha-1)^2$, we have
\[
\widehat{\text{BR}}_{g^{\text{LS}}}(\alpha) = 
\frac{1}{|\calI_T|}\sum_{j} \alpha(\widetilde{X}_j)^2 - \frac{2}{|\calI_S|}\sum_{i} \alpha(X_i),
\]
which yields the classical LSIF used in \citet{Kanamori2009aleastsquares}. While \citet{Chernozhukov2025automaticdebiased} proposes Riesz regression for covariate shift adaptation, their Riesz regression problem is the same as LSIF, and their covariate shift adaptation method basically coincides with \citet{Kanamori2009aleastsquares} except for the use of regression functions. In parallel, \citet{Kato2024doubledebiasedcovariateshift} also proposes a doubly robust form for covariate shift adaptation by extending \citet{Kanamori2009aleastsquares}.

\paragraph{KL divergence.}
For $g^{\text{KL}}(t)=t\log t - t$, we have
\[
\widehat{\text{BR}}_{g^{\text{KL}}}(\alpha)
=
\frac{1}{|\calI_T|}\sum_{j}\big\{\alpha(\widetilde{X}_j)\log \alpha(\widetilde{X}_j) - \alpha(\widetilde{X}_j)\big\}
-
\frac{1}{|\calI_S|}\sum_{i}\log \alpha(X_i),
\]
which is the standard UKL risk used by KLIEP–style procedures.

\paragraph{Power–divergence family (robust Bregman).}
For $g^{\text{PD}}_b(\alpha)=\bigp{\alpha^{1+b}-\alpha} / b$ with $b>0$, we obtain a continuum between Pearson–type ($b=1$) and KL–type ($b\to 0$) risks. Power–divergence choices trade efficiency and robustness: larger $b$ damp numerator outliers and can stabilize small–sample ratio learning, with convexity for $0 < b\le 1$.

\section{Automatic Covariate Balancing}
\label{sec:covariatebalancingconsistency}
Under specific choices of Riesz regression models and Bregman divergence, we can automatically guarantee the covariate balancing property. The key tool is the duality between Bregman divergence and covariate balancing methods. 

\subsection{Linear Models} 
Consider the linear model
\[\alpha_\beta(X) = \Phi(X)^\top \beta,\]
where $\Phi \colon \{1, 0\} \times \calZ \to \bbR^p$ is a basis function.  
For this model, using the squared loss (Riesz regression) automatically attains covariate balancing, as discussed in \citet{BrunsSmith2025augmentedbalancing}. 

Specifically, under linear models, from the duality, this MSE minimization problem is equivalent to solving
\begin{align*}
    &\min_{\alpha \in \bbR^n} \|\alpha\|^2_2\\
    &\text{s.t.}\ \ \sum^n_{i= 1}\alpha_i \Phi((D_i, Z_i)) - \p{\sum^n_{i=1}\Bigp{\Phi((1, Z_i))  - \Phi((0, Z_i))}} = \bm{0}_p,
\end{align*}
where $\bm{0}_p$ is the $p$-dimensional zero vector. 
This optimization problem matches that used to obtain stable weights \citep{Zubizarreta2015stableweights}.

It enforces the covariate balancing condition
\begin{align*}
   \sum^n_{i= 1}\widehat{\alpha}_i \Phi((D_i, Z_i)) - \p{\sum^n_{i=1}\Bigp{\Phi((1, Z_i))  - \Phi((0, Z_i))}} = \bm{0}_p,
\end{align*}
where $\widehat{\alpha}_i = \Phi(X_i)^\top \widehat{\beta}$.

The advantage of the use of linear models is that we can write the whole ATE estimation with a single linear models, as shown by \citet{BrunsSmith2025augmentedbalancing}. 

\subsection{Logistic Models}
We can model the Riesz representer via modeling the propensity score as
\[\alpha_\beta(X) = \mathbbm{1}[D = 1]r_\beta((1, Z)) - \mathbbm{1}[D = 0]r_\beta((0, Z)),\]
where 
\begin{align*}
    &r_\beta((1, Z)) = \frac{1}{\pi_{\beta}(X)},\quad r_\beta((0, Z)) = \frac{1}{1 - \pi_{\beta}(X)},\\
    &\pi_{\beta}(X) \coloneqq \frac{1}{1 + \exp\bigp{-\beta^\top \Phi(Z)}},
\end{align*}
and $\Phi \colon \calZ \to \bbR^p$ is a basis function. Note that we do not include $D$ unlike the basis function in linear models. 
For this model, if we use the KL-divergence–flavored convex function defined in Section~\ref{sec:empbalancing}, we automatically attain covariate balancing, as discussed in \citet{Zhao2019covariatebalancing}.

Define
\begin{align*}
    \widehat{\beta} \coloneqq & \argmin_{\beta}\frac{1}{n}\sum^n_{i=1}\Biggp{\mathbbm{1}[D_i = 1]\p{ - \log \p{\frac{1}{r_\beta((1, Z_i)) - 1}} + r_\beta((1, Z_i))}\\
    &\qquad\qquad\qquad\qquad + \mathbbm{1}[D_i = 0]\p{ - \log \p{\frac{1}{r_\beta((0, Z_i)) - 1}} + r_\beta((0, Z_i))}},
\end{align*}
and denote $r_{\widehat{\beta}}$ by $\widehat{r}$.

Specifically, under logistic models, from the duality, the KL divergence-flavored loss is equivalent to solving
\begin{align*}
    &\min_{w \in (1, \infty)^n} \sum^n_{i=1}(w_i - 1)\log (w_i - 1)\\
    &\text{s.t.}\ \ \p{\sum^n_{i= 1}\Bigp{\mathbbm{1}[D_i = 1]w_i \Phi((1, Z_i)) - \mathbbm{1}[D_i = 0]w_i \Phi((0, Z_i))}} = \bm{0}_p.
\end{align*}
This optimization problem matches that used in entropy balancing \citep{Hainmueller2012entropybalancing}.

As a result, we obtain 
\[\sum^n_{i= 1}\Bigp{\mathbbm{1}[D_i = 1]\widehat{w}_i \Phi((1, Z_i)) - \mathbbm{1}[D_i = 0]\widehat{w}_i \Phi((0, Z_i))} = \bm{0}_p,\]
where $\widehat{w}_i = \begin{cases}
    \widehat{r}((1, Z_i)) & \text{if}\ D_i = 1\\
    \widehat{r}((0, Z_i)) & \text{if}\ D_i = 0
\end{cases}.$

This model has the advantage that we can use a basis function $\Phi(Z)$ independent of $D$. Moreover, it naturally achieves covariate balance in the sense that the covariate distributions match between the treated and control groups. Additionally, we can automatically impose nonnegativity on $\alpha(1,Z)$ and $\alpha(0,Z)$, which can be violated in linear models.

\section{Conclusion}
This study proposed \emph{Direct Debiased Machine Learning} (DDML), a unified framework that targets the oracle Neyman score and estimates the Riesz representer via Bregman divergence minimization. DDML integrates and generalizes several lines of work in semiparametric efficiency and modern causal inference. In particular, Neyman targeted estimation provides a systematic route to estimate nuisance parameters by directly minimizing the estimation error between the oracle score and its plug-in counterpart, while generalized Riesz regression offers a flexible estimator of the Riesz representer that encompasses squared-loss Riesz regression, tailored-loss covariate balancing, and direct density-ratio estimation as special cases.

Conceptually, the framework connects strands that are often treated separately. We show that Riesz regression, covariate balancing propensity scores and empirical balancing, and least squares and KL type density-ratio procedures all arise from the same Bregman divergence minimization once the relevant score and representer are specified. This perspective explains when ostensibly different estimators coincide and when they differ, and it guides principled choices of losses and model classes for a given target parameter. Methodologically, the Bregman divergence shows that, by changing the convex function, we obtain objective functions that include Riesz regression and KL divergence minimization, enabling objectives that enforce automatic covariate balancing. 

In summary, DDML provides a single, coherent blueprint for constructing efficient, robust, and practically effective estimators for parameters defined through regression functionals. By casting nuisance estimation as Neyman targeted estimation and Riesz learning as Bregman divergence minimization, the framework unifies existing approaches and offers new tools that improve both theoretical guarantees and empirical performance. 

\bibliography{arXiv2.bb}

\bibliographystyle{tmlr}

\end{document}